\documentclass[
	a4paper, 
	10pt, 
	unnumberedsections, 
	twoside, 
]{LTJournalArticle}

\usepackage{amssymb}
\usepackage{natbib}
\setcitestyle{numbers,sort&compress}
\usepackage{cite}
\usepackage{amsmath,amssymb,amsfonts}
\usepackage{algorithmic}
\usepackage{graphicx}
\usepackage{url}
\urldef{\urlA}\url{https://github.com/dorsal-lab/tracecompass_nodejs_support} 
\urldef{\urlB}\url{https://github.com/Joker666/nodejs-memory-leak} 
\urldef{\urlC}\url{https://github.com/eclipse-theia/theia/issues/9514}
\urldef{\urlD}\url{https://github.com/acmeair/acmeair-nodejs}
\usepackage{subfig}
\usepackage{multirow}
\usepackage{listings}
\usepackage{xcolor}
\definecolor{codegreen}{rgb}{0,0.6,0}
\definecolor{codegray}{rgb}{0.5,0.5,0.5}
\definecolor{codepurple}{rgb}{0.58,0,0.82}
\definecolor{backcolour}{rgb}{0.95,0.95,0.92}
\newcommand{\infix}{\gg\hspace{-0.25em}=}

\usepackage{xcolor}
\definecolor{codegreen}{rgb}{0,0.6,0}
\definecolor{codegray}{rgb}{0.5,0.5,0.5}
\definecolor{codepurple}{rgb}{0.58,0,0.82}
\definecolor{backcolour}{rgb}{0.95,0.95,0.92}
\lstdefinestyle{mystyle}{
    backgroundcolor=\color{backcolour},   
    commentstyle=\color{codegreen},
    keywordstyle=\color{magenta},
    numberstyle=\tiny\color{codegray},
    stringstyle=\color{codepurple},
    basicstyle=\ttfamily\footnotesize,
    breakatwhitespace=false,         
    breaklines=true,                 
    captionpos=b,                    
    keepspaces=true,                 
    numbers=left,                    
    numbersep=5pt,                  
    showspaces=false,                
    showstringspaces=false,
    showtabs=false,                  
    tabsize=2
}
\runninghead{Node Compass: Multilevel Tracing and Debugging of Request Executions in JavaScript-Based Web-Servers} 

\title{Node Compass: Multilevel Tracing and Debugging of Request Executions in JavaScript-Based Web-Servers} 

\author{%
	Herve M. Kabamba\textsuperscript{1}, Matthew Khouzam\textsuperscript{2} and Michel Dagenais\textsuperscript{1}\thanks{Corresponding author: \href{mailto:herve.kamba-mbikayi@polymtl.ca}{herve.kamba-mbikayi@polymtl.ca}\\ \textbf{Received:} November 17, 2023 }
}
\date{\footnotesize\textsuperscript{\textbf{1}}Computer and Software Engineering Department, Polytechnique Montréal\\ \textsuperscript{\textbf{2}}Ericsson Canada\\ }
\renewcommand{\maketitlehookd}{%
	\begin{abstract}
		\noindent Adequate consideration is crucial to ensure that services in a distributed application context are running satisfactorily with the resources available. Due to the asynchronous nature of tasks and the need to work with multiple layers that deliver coordinated results in a single-threaded context, analysing performance is a challenging task in event-loop-based systems. The existing performance analysis methods for environments such as \texttt{Node.js} rely on higher-level instrumentation but lack precision, as they can’t capture the relevant underlying application flow. 

As a solution, we propose a streamlined method for recovering the asynchronous execution path of requests called the Nested Bounded Context Algorithm (NBCA). The proposed technique tracks the application execution flow through multiple layers and showcases it on an interactive interface for further assessment. Furthermore, we introduce the vertical span concept. This representation of a span as a multidimensional object (horizontal and vertical) with a start and end of execution, along with its sub-layers and triggered operations, enables the granular identification and diagnosis of performance issues. We proposed a new technique called the Bounded Context Tracking Algorithm (BCTA) for event matching and request reassembling in a multi-layer trace . The two techniques allow aligning the executions of the request in a tree-based data structure for developed visualisations.
These visualisations permit performance debugging of complex performance issues in Node.js.
	\end{abstract}
}

\begin{document}

\maketitle 


\section{Introduction}
\label{sec:intro}
The rapid growth of distributed applications, combined with recent advancements in cloud-based application development tools and technologies, have created a significant need for performance analysis and optimization of those applications. This These advancements come with a cost in managing performance.The complexity of debugging is one of the primary issues that arise in distributed systems, as explained by \citet{Maximilian2020}. Indeed, distributed applications often rely on many abstraction layers to ease communication, deployment, scaling, performance, and resilience. 

Furthermore, the separation of services into independent components, the containerization, the choice of communication protocols amongst micro-services, and various additional factors complicate the monitoring and debugging of these systems. Performance analysis remains a significant challenge, generating considerable interest in the topic. Although numerous solutions and tools \citep{desnoyers2006lttng, aguilera2003performance, clinic, compass}, are available, the environment in which the application runs, and the technologies employed, generally dictate the methodology for evaluating the performance of these applications.

Distributed applications typically use distributed tracers, to capture data on the application specific interactions, at the expense of instrumentation. A collector then aggregates the distributed traces to make them available for the user interface modules~\citep{tigrannajaryan}. For instance, the collector can gather and hierarchically aggregate the information about the request life-cycle, and the correlations between the sub-request interactions with other system components. For \texttt{JavaScript} applications, particularly those written in the \texttt{Node.js} environment, the simple use of distributed tracers to monitor and pinpoint potential performance issues creates a significant concern. The \texttt{Node.js} infrastructure comprises multiple layers, each performing a distinct function in processing and executing various tasks.

Moreover, the single-threaded nature of \texttt{Node.js} requires the main thread to instantiate an event-loop (EL), which orchestrates the execution of tasks by passing through several phases~\citep{Anastasia2020}, with the main thread responsible for callbacks invocation on the one hand. On the other hand, it dispatches lengthy (possibly blocking) tasks to a thread pool to expedite their execution and avoid blocking while waiting for lengthy operations to complete. This intricacy is all the more noteworthy, because relying solely on a distributed tracer to instrument higher-level functions does not allow for an objective assessment of application performance in this context. Such environments require a greater level of granularity to precisely determine the underlying cause of any performance issue, since submitting asynchronous requests to the lower layer does not guarantee serial and instantaneous execution. The latter must first be queued with other concurrent tasks and awaits orchestration by the EL.

The time delay of a single request, or its nested requests, is not a reliable source of information for identifying the issue in this specific situation. For instance, slow code execution can keep the EL in a specific phase for an extended period of time, preventing it from moving to another phase to execute the many pending callbacks. In turn, the issue propagates to other pending requests, increasing the delay of other spans in the distributed trace. Since \texttt{Node.js} applications load and interact with many other modules, non-instrumented functions are unable to give performance metrics for diagnosing a specific performance issue, resulting in fault propagation to the pending requests.

Furthermore, the inner workings of \texttt{Node.js} prioritize a set of tasks en-queued into specific data structures, and ensure that they are exhausted before moving to the next phase. If the executing code is slow in this context, the transition latency between phases will increase, and the application performance will degrade. Consequently, the latency of pending tasks will increase rapidly, even though they are not the root cause of the problem.

This paper presents an efficient method for analyzing the performance of \texttt{Node.js} applications based on metrics monitoring and root cause analysis. We use the \textit{Nested Bounded Context Algorithm} (NBCA), that exposes the request pathways, and the \textit{Bounded Context Tracking Algorithm} (BCTA), that reconstructs the request vertical execution sequences into a span, allowing for fine-grained diagnosis and pinpointing of performance issues.

To the best of our knowledge, there is no pre-existing
efficient technique that allows a global performance analysis with such level of granularity for \texttt{Node.js} applications.

The following are the primary contributions of this paper: (1) We present a novel trace span representation model, named the \textit{Vertical Span Model} (VSM), designed to describe the operations of a single request across several layers and sub-layers, along with its metadata. (2) We present a fine-grained method for analyzing the performance of \texttt{Node.js} applications using kernel and user traces. We employ a four-layer paradigm to collect data at the \texttt{JavaScript} interpreter, the \texttt{virtual machine},  the \texttt{libuv}, and \texttt{kernel} levels. (3) We introduce a new crucial performance metric named the \textit{Atomic Task Time Latency} (ATTL) that measures the internal atomic tasks execution time. (4) We build a state system model that captures the traces events attributes and metadata. (5) A multi-level traces events correlation technique is presented. (6) We demonstrate the relevance of our tool and the effectiveness of our technique with realistic use cases. (7) We implemented many interactive views for diagnosing system issues.

The remainder of the paper is organized as follows: Section~\ref{sec:related_work} discusses related work about the subject. Section~\ref{sec:methodology} presents the proposed solution.  Section~\ref{sec:use-cases} presents some use cases to show our work pertinence and relevance. In Section~\ref{sec:evaluation}, we discuss the results obtained while in Section~\ref{sec:conclusion} a conclusion on the work is drawn. 

\section{Related Work}
\label{sec:related_work}

The rapid transition of programs from monolithic architectures to distributed systems, particularly in Cloud Computing environments, has created a significant need for performance management~\citep{appdynamics}. Performance management, in this case, is analyzing the behavior of the program in order to determine if it deviates from its typical behavior and, if so, to identify the root cause of the problem and thereby resolve it. Various works related to \texttt{Node.js} have been proposed in the literature. The run-time monitoring was applied to \texttt{Node.js} by \citet{ancona2018towards} to verify the correctness of an asynchronous nested callback in the context of the Internet of Things (IoT). They employed the parametric trace expression to monitor running applications developed in \texttt{Node.js} and Node-KED, which is a tool built on top of \texttt{Node.js} for flow-based IoT programming. The correctness of the application is verified at run-time through application tracing mechanisms, to check the compliance of the system behavior with the expected behavior defined by a specification formalism. 

\citet{chang2019detecting} investigated the process for detecting atomicity violations in event groups while studying the performance of \texttt{Node.js}. A task is a collection of discrete, deterministic events. They addressed event-race conditions within the context of many events. Tracing was utilized to gather information about the execution of the analyzed system, in order to predict and discover atomicity violations within a set of events. The identification of event pairings was based on the Happens-before relationship present in the execution trace. Atomicity violations were identified using predefined patterns.

Different approaches to the performance analysis of asynchronous \texttt{JavaScript} applications have been presented in the literature. Some approaches have been confined to detecting race conditions using client-side web application code. Predictive approaches are based on the recording of a part of the application execution by performing an analysis of the different memory accesses and the causal dependencies between the executions of the event handlers, and the data of the various execution scenarios, which form the basis for the generation of race conditions.

\citet{jensen2011modeling} introduced a browser-based data flow analysis for \texttt{JavaScript} applications named TAJS. The authors focused on representing the Document Object Model (DOM) instead of considering the events in depth. The approach assumes at the same time that any event generated by the browser can be executed at any time but cannot be executed in the context of separation of data flow, based on which listeners are registered. GATEKEEPER was introduced by \citet{guarnieri2009gatekeeper} as a tool for enforcing specific \texttt{JavaScript} widget security policies. In particular, GATEKEEPER allows for the detection of unauthorized data flows. The concurrency problem on the client side was addressed by \citet{hong2014detecting} who presented WAVE, which is a framework for detecting concurrency errors. It is based on a model emanating from the execution. Test cases are generated and allow to swap the order of operations on the original execution flow. Any exception generated as a result of the test, or any result contrary to the original execution flow, constitutes an error reported by the system.

Under \texttt{Node.js}, few server-side race detection approaches have been utilized. NodeAV~\citep{chang2019detecting}, is only capable of detecting atomicity violations. \citet{davis2017node} applied fuzzing to the \texttt{Node.js} scheduler for race detection. Internal \texttt{Node.js} queues were used to randomize the association of inputs in order to maximize the probability of discovering faulty executions during testing. This technique enables the identification of several race conditions but is not applicable to issues that manifest during the delay of specific events. This method also makes it challenging to identify the underlying cause of the performance issue. With NodeRacer~\citep{endo2020noderacer}, investigations on alternate schedules by selectively delaying events, without affecting \texttt{Node.js} core queues, was conducted. The work was predominantly influenced by \citet{adamsen2017repairing} and \citet{adamsen2018practical}. The analysis is based on the formalism of the happens-before relationship, which enables the identification of causal relationships between events. The initial running of the application enables the inference of the happens-before relationship based on either a regression test or manual interaction with the application.

The network performance of \texttt{Node.js} was investigated by \citet{nkenyereye2016performance} as they evaluated the performance of a Healthcare Hub Server for a Remote Monitoring System. For this latter, emphasis was placed on concurrency, because \texttt{Node.js} contains all the necessary technology to effectively manage its issues. Throughput and response time were measured using \texttt{Apache Jmeter}\citep{jmeter}. \citet{madsen2015static} used static analysis on \texttt{Node.js} to identify problems in the events processing. According to their observation, ordinary call graphs are unsuitable because they do not accurately depict the control flow implied by the registration of event listeners and the emission of events. Therefore, they proposed an extension of the Call Graph and implemented some analyses using the static analysis tool RADAR. 

Most \texttt{Node.js} performance analysis tools tend to focus on specific areas of action. As mentioned earlier, various approaches are sometimes limited to detecting race conditions, atomicity violations, or analyzing concurrent flow.

These studies adequately demonstrate the complexity of tracking event orchestration in \texttt{Node.js} due to its nondeterministic nature in task execution. They also underscore the need for tools capable of detecting issues in this context. However, performance problems in \texttt{Node.js} are generally addressed at the highest level, typically in JavaScript. 
Bottlenecks can occur in the lower layers of the architecture and remain completely hidden from current analysis tools. More detailed techniques are required to identify performance issues at the internal layers that orchestrate and execute them.

Furthermore, the few tools that address specific problems in \texttt{Node.js} use the Async Hooks API to instrument applications for collecting execution traces. 
It is well-known that activating this API imposes a significant overhead on systems. Node Compass introduces a low-overhead approach to data collection and offers a multilevel approach to system execution traces. To the best of our knowledge, there is no existing research on multilayer tracing in \texttt{Node.js}.

Therefore, it is imperative to develop more comprehensive and efficient tools for performance analysis in \texttt{Node.js} that can delve into the intricacies of its internal layers while minimizing overhead, as the current state of the art falls short in addressing these critical aspects.
\section{Proposed solution}
\label{sec:methodology}

\subsection{The Vertical Span Representation Model}
\label{sec:vertical_span_representation}

The vast majority of \texttt{Node.js} performance analysis approaches rely on distributed tracing to collect the necessary data. This information, combined in a single trace, enables the observation of the request execution latency in the form of spans. The metric is the duration of the execution of each request component, and it contains very high-level information. The \texttt{Node.js} architecture comprises a collection of components that may be classified into three distinct layers. The top layer is the \texttt{JavaScript} interpreter, followed by the internal layers, namely the \texttt{V8} engine, \texttt{libuv}, and the operating system kernel. Distributed tracers Instrumentation present a significant challenge as they expose only symptoms of the poorly performing application.

In general, the instrumentation will be performed at the upper layer level, which can only provide an overview of a performance issue, without allowing it to be precisely pinpointed, due to the complexity of the environment. For instance, calling a \texttt{JavaScript} file \texttt{read} function may incur considerable lag, suggesting a performance issue. However, this information cannot be used to establish the root cause of a problem. Indeed, the latency could be caused by a blocking state in the EL or by the exhaustion of the threads in the pool, to mention a few. As a result, it is crucial to employ alternative performance analysis techniques, offering a higher degree of granularity, in order to isolate the cause of the problem. The VSM provides the vertical representation of a span. In the latter model, the various sub-layers of the system involved in the execution of the request are used to reconstruct the sequences of execution and granularly display the span for a precise pinpointing of the performance problem.

\begin{figure*}[ht]
\centering{\includegraphics[width=7cm,height=5cm]{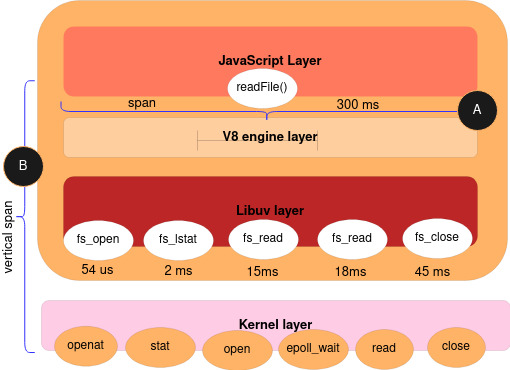}} 
\caption{Vertical Span Model (VSM) representation. The label “A” depicts a span as it appears in distributed
tracers. The label “B” depicts a vertical span representing the flow of the request sequences in all layers.}
\label{fig:vspan}
\end{figure*}

The span in Figure~\ref{fig:vspan} was obtained by instrumenting a \texttt{JavaScript} \texttt{readFile} function call. The task execution time (300 ms) is denoted by the span-shaped label ``A''. Label ``B'' shows the VSM, which depicts the synchronous and asynchronous operations of the different sub-layers that make up the span. As seen in Figure~\ref{fig:vspan}, calling the \texttt{readFile} method, in the higher layer, triggers a set of asynchronous function calls that the \texttt{libuv} layer will execute. At this level, the \texttt{fs\_read} function may be transparently invoked several times. The execution of I/O functions triggers related system calls. Since function calls are not sequentially executed in the different layers, but are instead coordinated by the EL, it is challenging to figure out the root cause of the performance problem. The VSM enables the definitive identification of the nature of issues at hand.

\subsection{Asynchronous Operations in \texttt{Node.js}}
\label{sec:asynchronous_operations}
The execution of asynchronous operations is a core feature of \texttt{Node.js} that allows the execution of other instructions without waiting for blocking and slow tasks. To do this, an internal orchestration system in \texttt{Node.js} allows the execution of tasks, and the storage of various related references. An asynchronous operation in \texttt{Node.js} embeds a callback function that is invoked at the end of its execution. \\\\
\textbf{Definition: } an atomic operation is the smallest component of a high-level function that triggers system calls or that can be executed directly by the Node.js runtime.\\\\

Formally, the asynchronous operation can be seen as a monad~\citet{monad}. In Figure~\ref{fig:vspan}, the \texttt{Node.js} \texttt{readFile} function triggers a set of I/O function calls at the \texttt{libuv} layer, which can be interpreted as follows: the \texttt{fs\_open} function is called while the \texttt{fs\_stat} is passed to it as a callback. The invocation of the callback function embeds the function \texttt{fs\_read} as a callback, and likewise, the latter is invoked with the callback function \texttt{fs\_close} passed to it as an argument.

This sequence of asynchronous operations can be represented formally as a monad. Consider $R$ as the result of executing the I/O function $f$ at the  \texttt{JavaScript} interpreter layer. By projecting the execution of the latter on the bottom \texttt{libuv} layer, as shown in Figure~\ref{fig:vspan}, the result $R^{'}$ obtained there must be equivalent to $R$. Based on \citet{janin2020equational} work, we can derive the result $R$ by:
\begin{equation}
	R=f_{1} \infix f_{2} \infix f_{3} \infix f_{4} \infix f_{5}
	\label{eq:1}
\end{equation}
\noindent where $\space R\equiv R^{'}$.

The chained execution of these I/O functions makes a monad reference memory that the parent function can use to get the chain result. In this case, the monad reference precludes the execution of higher-level functions. This reference points to an active task. Under \texttt{Node.js}, it will often be a memory address used to access the value returned by the referenced task.

Let $m$ be a monad of type $T$, referenced by a pointer to its forked action $P_{m}$, as modeled in \citep{janin2020equational}.

\begin{equation}
	P_{m}::*\to *
\end{equation}
\begin{equation}
	\texttt{fork}::m\:  x\to m(P_{m}\: x)
\end{equation}
\begin{equation}
	\texttt{read}::P_{m}\:  x\to m\: x
\end{equation}

$P_{m}$ is a monad reference bound to an active action of type $m\ x$, \texttt{fork} $m$ is the activator of the execution of the monadic action $x$ and returns a reference bound to that action, and \texttt{read} $p$ is an action that waits for and accesses the memory reference, to return the value returned by the running action bound by the monad reference $p$.

In Equation~\ref{eq:1}, the chained execution of functions requires the passing of arguments of objects references bound to these functions, and to events that provide information on the state of the function execution. The triggering of these events allows access to the addresses containing the returned values. A value indicating the \texttt{ATTL} is attached to each atomic function $f1, \dots, f5$. The execution time $\mathbb{T}$ can therefore be obtained by:

\begin{equation}
	\mathbb{T} = \Sigma_{i=1}^{n} t(f_{i})
\end{equation}
\begin{equation}
	\mathbb{L} = \mathbb{T} + \Sigma_{i=1}^{n-1} \Delta(t_{i,i+1})
\end{equation}

The overhead latency $\mathbb{L}$ of the \texttt{JavaScript} function is computed by summing the waiting times $\Delta(t_{i-1,i})$, for each function $f_{i}$ to be executed from the completed execution of the predecessor $f_{i-1}$. In this case, $\Sigma_{i=1}^{n-1} \Delta(t_{i-1,i})$, constitutes the overhead of the request execution, and a threshold must be fixed to its value to declare a performance issue. The VSM allows the visualization of each atomic operation, along with its latency throughout the involved layers. The reader is referred to the work of \citet{janin2020equational} and \citet{moggi1991notions} for additional information on monad functions.

\subsection{Task Transitions and Atomic States}
\label{sec:request_transition}

\begin{figure}[htb]
	\centering
	\includegraphics[width=0.6\linewidth]{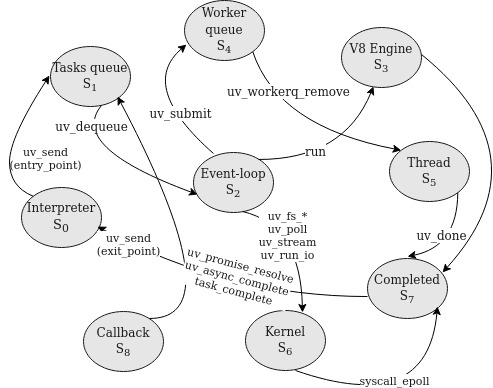}
	\caption{Request states during runtime. Nodes represent the state of tasks as seen by the Node.js runtime and edges represent the transitions labeled with the triggered events.}
	\label{fig:request_states}
\end{figure}

The interpreter is the entry point for a \texttt{JavaScript} function execution. This is the request $S_{0}$ state, as depicted in Figure~\ref{fig:request_states}. The latter is subsequently sent to the bottom layer, which queues it. Then it moves to state $S_{1}$. The transition between the two states is captured by the \texttt{uv\_send} event in the trace. At this point, the request is waiting to be removed from the queue in the order in which it was received. The EL conducts continual checks and executes the different queued jobs in a First In, First Out (FIFO) pattern. The task is then removed from the queue and again queued for the corresponding EL phase. Then it moves to state $S_{2}$. Function \texttt{uv\_dequeue} is used to instrument this transition.

The event loop will see the request and manage its execution by the \texttt{Node.js} runtime, as it moves through different stages. At this level, three paths are possible: If the request is a \texttt{Node.js} operation, the runtime executes it instantly. It then transitions to state $S_{3}$, which is instrumented by the \texttt{run} event. If the task involves an I/O operation, it is pushed into a worker queue and transitions to $S_{4}$ state. The latter, \texttt{uv\_submit}, is instrumented. It will then be pulled from the worker queue by a thread from the pool, and the associated I/O operations will be performed. In this case, it transitions to state $S_{5}$. Function \texttt{uv\_workerq\_remove} is used to instrument the transition. The third possible path for the task is to move to state $S_{6}$. In this case, the operating system runs the task and notifies the upper layer through the built-in asynchronous communication mechanisms.

Functions \texttt{uv\_fs\_*} and \texttt{uv\_socketRead} are used for many input and output operations. For example, \texttt{uv\_socketRead} is used to read from a socket. The $S_{7}$ state is instrumented by \texttt{uv\_done} and is used to indicate the completion of a task. This event returns an atomic operation unique ID. Recall that an atomic operation is one part of the chain of execution of a request made by the interpreter, but there are many more. The result can be sent back directly to the interpreter from $S_{7}$ state. The transition is instrumented by \texttt{uv\_send} as the exit point. It can also return to state $S_{1}$ from state $S_{7}$, if a monad reference was given to the superior function.

In the case of the \texttt{readFile} function, the \texttt{fs\_open} atomic task returns a file descriptor that is provided as an argument to subsequent atomic file access operations.

\subsection{System Architecture}
\label{sec:system_architecture}

\begin{figure}[ht]
\centering
\includegraphics[width=0.7\linewidth]{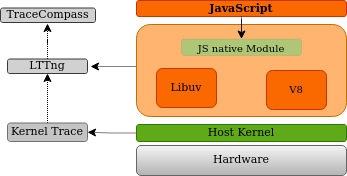} 
\caption{Architecture of the collection and analysis system.}
\label{fig:architecture3}
\end{figure}

The system architecture is shown in Figure~\ref{fig:architecture3}. The data is collected through tracepoints inserted into the system. The \texttt{LTTng} (Linux Tracing Toolkits Next Generation) tracer is used to perform the static instrumentation. A native module called \texttt{calculate} allows tracepoints injected into \texttt{JavaScript} functions to invoke the \texttt{LTTng} tracer libraries at the highest level. Static tracepoints are also added to the \texttt{libuv} layer to collect information on atomic tasks orchestration, and in the \texttt{V8} engine to collect information on the asynchronous resources life-cycle. Kernel traces are obtained with \texttt{LTTng}.

The system instrumentation allows two synchronized traces to be loaded into Trace Compass\citep{compass} as an experiment. In the latter, precise analyses and algorithms are run, metrics are extracted, and views are built to observe the system performance. Because of its low overhead on the system~\citep{desnoyers2006lttng}, the \texttt{LTTng} tracer was chosen from a set of tracers. \texttt{LTTng} is considered today as the fastest Linux tracer~\citep{10.1145/3158644}. An extension of the tool enables the analyses to be run for Windows \texttt{Node.js} versions.

The process for conducting the analysis is as follows:
\begin{enumerate}
	\item Start the tracer.
	\item Launch the concerned \texttt{Node.js} application.
	\item Stop the tracer on the host.
	\item Run the analysis.
	\item Use an interactive tool (Trace Compass) to visualize the results.
\end{enumerate}

\texttt{LTTng} generates and collects events, which are then loaded into Trace Compass. On the latter, an event analyzer was developed to handle our trace events. Trace Compass is a free and open-source program that analyzes traces and logs. Its extensibility enables the creation of views and graphs, as well as the extraction of metrics.

\subsection{Nested Asynchronous Operations Detection}
\label{sec:nested_asynchronous_operations_detection}

The execution of a \texttt{JavaScript} function consists of a group of atomic operations that happen across the system layers. The \texttt{Node.js} environment handles requests made at the interpreter level in a specific execution context. It is feasible to obtain information about the context $id$ by inserting tracepoints that are activated during the initialization of asynchronous resources, and immediately following their execution.

\begin{lstlisting}[language=Octave, label={ref:snipet1},caption={JavaScript instrumented code snipet. The \texttt{calculate} module is the \texttt{C++} addon that calls the LTTng functions activating the tracepoints.}]
app.get('/compute-with-promises', function computePromise(req, res) {
  const as=async_hooks.executionAsyncId();
  const ass= async_hooks.triggerAsyncId();
  calculate.send_event(as, ass, 'js_open_computePromise');
  const hash = readData();
  const updatePromise = () => new Promise((resolve) =>{
     calculate.send_event(async_hooks.executionAsyncId(), async_hooks.triggerAsyncId(), 'jsx_updatePromise');
     hash.update();     
     resolve(); });
  const loop = () => {   
      updatePromise().then(loop).then(calculate.send_event(as, ass, 'js_exit_computePromise'));
    } }
  loop();  });
\end{lstlisting}
\begin{figure*}[!ht]
\centering{\includegraphics[width=13.5cm,height=6cm]{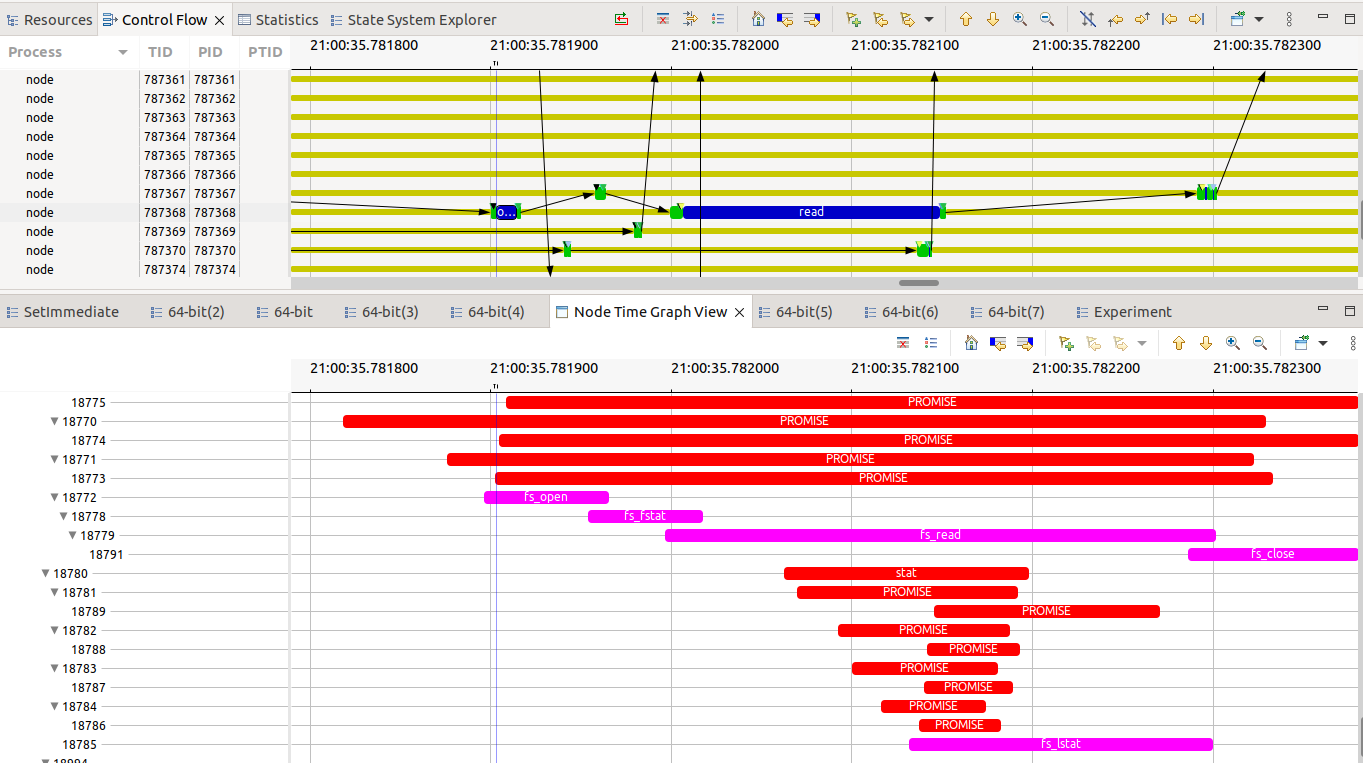}} 
\caption{Execution Flow and Nested Asynchronous Operations. The Control Flow perspective shows a timeline of the multi-layer events. At the bottom, the Node TimeGraph view depicts atomic operations flow in the middle and their executions contexts numbers in the left.}
\label{fig:flow_good1}
\end{figure*}

The \texttt{uv\_send} event is used to instrument \texttt{JavaScript} functions and to track the asynchronous objects life-cycle. In Listing \ref{ref:snipet1}, the \texttt{AsyncHooks} API is used along with the \texttt{calculate} module. The latter accepts an $id$, a \texttt{channel}, and a \texttt{method} as arguments. When an \texttt{HTTP} request is received, function \texttt{ComputePromise} is called. A tracepoint is then triggered embedding the information about the execution context $id$ as well as the $id$ of the resource that triggered the current asynchronous object. At line $4$, the function entry point is instrumented. Information about the request context identification number can be found in the $id$ field, while information about the $id$ of the resource that triggered the current context can be found in the $channel$ field. There is a field called \texttt{method} that has the name of the function that was instrumented. In this instance, it is \texttt{computePromise}. The $channel$ field can also contain the name of the asynchronous resource, like \texttt{promise} or \texttt{FSReqCallback}, among many others. It is obtained by getting the value of the \texttt{type} variable.

At the end of the function execution, \texttt{uv\_send} from the \texttt{calculate} module is used again to signal the completion of the request. It can be observed at line $11$ in the Listing \ref{ref:snipet1}. In this instance, the $id$ field includes the same unique identifier as the initial request, the $channel$ field is $null$, and the field \texttt{method} has the ``\texttt{js\_exit}'' value. In this case, it signifies the end of the \texttt{JavaScript} function execution. At line $6$ a new asynchronous resource is triggered creating a new execution context bound for the asynchronous resource. The latter is instrumented at line $7$. In this case, the execution context $id$ is new. The $id$ of the resource that triggered the new object, is the execution context $id$ of the \texttt{computePromise} function. We do not need to instrument the completion of asynchronous resources, since this will be obtained from the \texttt{V8} engine directly. The NCBA algorithm uses the \texttt{uv\_send} event to figure out the nested execution order of the atomic operations. Instrumentation of the \texttt{JavaScript} function entry and exit points is needed to get back the time it took to run.

However, the goal of the algorithm is to figure out the nested flow of operations. In response to the \texttt{uv\_send} event, the latter will look at the value of the \texttt{channel} field, which contains the $id$ of the resource execution context, and will query the State History Tree (SHT). The SHT is a tree-shaped data structure that efficiently stores and indexes the attributes and the states extracted from the trace. It will get the $id$ of the parent resource that created the current resource execution context. This SHT will be queried again using the obtained parent resource execution context $id$ to obtain the $quarkParent$, which is the value specifying the position in the SHT at which a new child entry will be created.

The algorithm will store the new entry position once more, and will update its status with the name of the method in the SHT. It will then monitor the ``\texttt{run}'' or ``\texttt{resolve}'' values in the \texttt{method} field of the \texttt{uv\_send} event, to later alter the entry status, initially set to $null$, based on the $id$. This signals the end of the asynchronous resource execution. As it traverses the trace, the algorithm repeats the same operations. Finally, the algorithm outputs an SHT comprising all the operations, in hierarchical order, along with their execution time.

In Figure~\ref{fig:flow_good1}, on the left panel of the Node TimeGraph view, the execution context IDs of the different atomic operations are identified and displayed in a hierarchical order, as computed by the algorithm. For instance, the \texttt{readFile} function, executed at the top layer, triggered the execution of the \texttt{fs\_open} function on the \texttt{18772} execution context, at the \texttt{libuv} layer. The latter triggered the execution of the \texttt{fs\_stat} function on the execution context \texttt{18778}, which again triggered the execution of the \texttt{fs\_read} function on the execution context \texttt{18779}, and finally the latter triggered the execution of the \texttt{fs\_close} function on the execution context \texttt{18791}. On the right panel, the computed nested execution flow can be seen. Some operations may trigger promises, and the latter may trigger atomic operations within their execution contexts. Exposing the execution flow allows a granular identification of the root cause of performance issues.

\subsection{Multi-layer Events Matching}
\label{sec:multi-layer_events_matching}
In the context of multi-layer event matching, our focus is directed towards two primary concepts: vertical profiling and calling-context profiling. The first has been investigated by \citet{hauswirth2004vertical} and \citet{hauswirth2005automating}. It entails profiling various metrics from various layers of the system stack. Analysis of application performance is complicated by the addition of multiple abstraction layers to the execution environment. The methodologies investigated and suggested for vertical profiling predominantly rely on multi-layered systems that enable multi-threading. Nevertheless, these methodologies are not congruent with event-driven systems such as Node.js. 
\begin{figure}[!htb]
	\centering
	\includegraphics[width=0.6\linewidth]{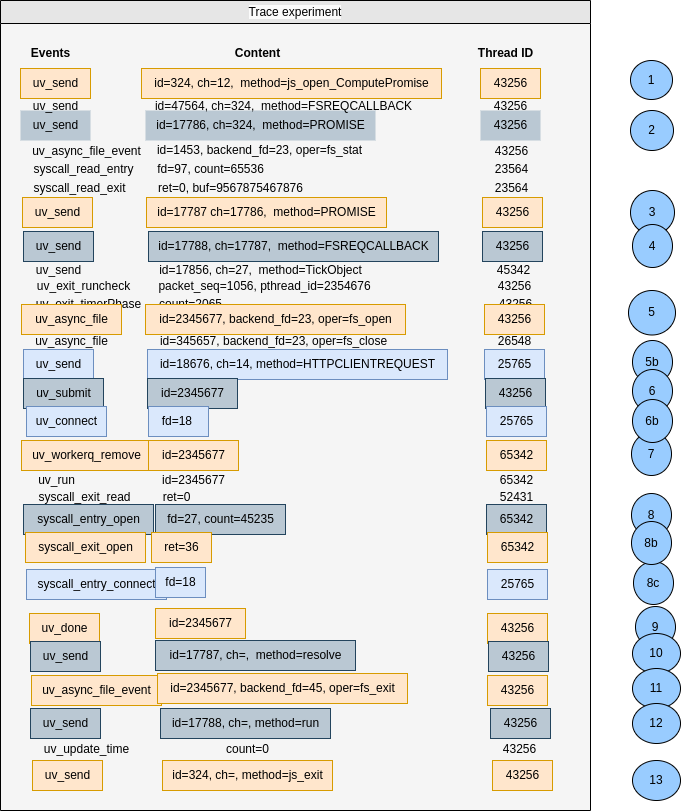}
	\caption{Multi-layer trace. Event from different layers are aggregated and loaded as an experiment. For clarity some contents, fields and timestamps have been omitted}
	\label{fig:trace}
\end{figure}

The operating system views these systems as black-box devices whose operations are not executable in a deterministic manner. An EL orchestrating their execution introduces a horizontal overhead. This EL also adds an extra layer of abstraction, complicating any performance analysis. Calling-context profiling \citet{ausiello2012k} and \citet{moret2009cccp}, uses a tree data structure to provide dynamic metrics for each calling-context. 

We propose a new approach that uses a state system created from trace attributes data. It employs a state history tree (SHT), a data structure that is optimised during the construction of our model while traversing the trace, to store the attributes extracted from the latter. The platform for modelling and developing such data structures is provided by Trace Compass. Their optimisation enables access to the model with a logarithmic complexity. Taking into account the two proposed concepts, we developed a profile of the analysed system based on the upstream trace. This is a crucial and difficult step, particularly considering there is virtually no literature on the subject. 

It is obvious that building a calling-context profile within the same layer is a straightforward approach. However, in such a setting, it is quite difficult to correlate the same information, arriving from various layers, while also connecting them to the functions from higher layers. To the best of our knowledge, there is no documented method for accomplishing this using Node.js. Our Node Compass approach instruments, aggregates, and constructs a model utilising the state system and SHT data structures. The latter is constructed by looking for patterns in the multi-layer trace that enable the reconstruction, of the vertical and horizontal execution sequences, of the system under inspection. The end result is a visual and interactive instrument that enables the user to identify problems.

Basically, two types of traces are obtained from the system instrumentation, as depicted in Figure~\ref{fig:trace}: userspace and kernel-space traces. The BCTA is used to look for patterns in the multi-layer traces and allows the correlation of events. In Figure~\ref{fig:request_states}, after being submitted in the upper layer, the request moves to state $S_{2}$, where the intermediary layers manage its execution. It can be observed that the request can take the following state paths: $S_{0} \to S_{1} \to S_{2} \to  S_{3} \to  S_{7} \to S_{0}$, $S_{0} \to S_{1} \to S_{2} \to S_{6} \to S_{7} \to S_{0}$, and $S_{0} \to S_{1} \to S_{2} \to S_{4} \to S_{5} \to S_{7} \to S_{0}$. In the first case, it is a \texttt{Node.js} non-computational intensive task. Il is executed right away by the runtime. In this case, the higher-layer function is instrumented by the \texttt{uv\_send} event, which identifies the entry and exit points of the function  Figure~\ref{fig:trace} (labels $1$ and $13$).

In the second case, the task execution is delegated to the operating system (OS). The interpreter, the kernel, and the \texttt{libuv} layers will all be involved in the event matching process, Figure~\ref{fig:trace} (labels $5b$, $6b$ and $8c$). At this level, events that cause the OS to run system calls are captured. These latter are sequentially preceded by a resource initialization process in the \texttt{virtual machine}, Figure~\ref{fig:trace}, label $5b$. Directly inserted tracepoints enable to obtain the information on the execution context, and the name of the resource being executed. The BCTA will first look for the closest \texttt{uv\_send} event that precedes the \texttt{libuv} atomic task (label $6b$) based on its thread ID, since both events are sequential. This \texttt{uv\_send} event, obtained from the \texttt{VM} instrumentation, provides the context data required to match the appropriate \texttt{JavaScript} function, instrumented with the  \texttt{AsyncHooks} API. Consequently, it is possible to reconnect operations from these three layers, as Figure~\ref{fig:trace} shows.
\begin{figure}[!htb]
	\centering
	\includegraphics[width=0.8\linewidth]{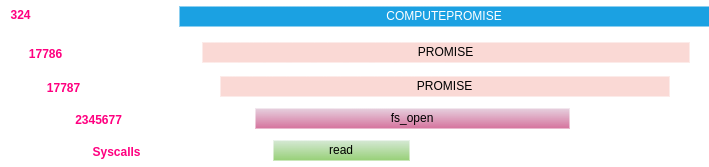}
	\caption{Resulting vertical span from the NBCA and BCTA execution of the trace in Figure~\ref{fig:trace}}
	\label{fig:exspan}
\end{figure}
The upper layer and the \texttt{libuv} events are linked by creating a new child entry at the position of the \texttt{JavaScript} initial function in the SHT. To match the kernel events, the BCTA tracks all system calls executed by the same thread. It creates child entries at the corresponding \texttt{libuv} atomic task position in the SHT for each one, as long as the request execution is not completed. This information is obtained by watching the \texttt{uv\_send} events, which signal the \texttt{JavaScript} functions exit point. The algorithm uses this information as a constraint, before searching for patterns. These execution bounds are named the \textit{Bounded Execution Context} (BEC).

For the third case, consider the execution of a JavaScript function that opens a file. The \texttt{libuv} layer utilises a thread pool, which simulates the asynchronous execution for blocking processes. Thus, it can continue its execution and move through the various phases, invoking the callbacks that are awaiting execution. In the trace presented in Figure~\ref{fig:trace}, the \texttt{ComputePromise} function that opens a file is called. The instrumentation of the beginning of the function execution triggered the event uv\_send with the id=$324$. The BCTA will track the resource id=$324$ and reach label $2$. A \texttt{promise} is created from the execution context of the higher level function \texttt{ComputePromise}. The BCTA will keep tracking the id which is now $17786$ and reach label $3$, where a new promise has been created from the execution context of the preceding \texttt{promise}. It will then track the id=$17787$ which captures the beginning of a file system operation. Constrained by the BEC, the BCTA relies on the events \texttt{uv\_async\_file}, \texttt{uv\_submit}, and \texttt{uv\_workerq\_remove}, which respectively indicate, the beginning of a file system atomic task, the queuing of the task by the main thread executing the EL, and its removal from the queue by an available thread in the pool. These events are obtained from the \texttt{libuv} instrumentation. It can be observed in Figure~\ref{fig:trace}, label $7$, how a  thread from the pool is removing the task from the queue (different thread ID). The algorithm seeks for the \texttt{uv\_send} event from the \texttt{VM} instrumentation that immediately precedes the \texttt{uv\_async\_file} event, as depicted in Figure~\ref{fig:trace}, label 4. In this way, the execution context information is obtained by the algorithm.

These events provide the BCTA with sufficient information to establish a connection between the task being executed by the thread and the main process that runs the EL, as well as the initial \texttt{JavaScript} function. The atomic operation $id$ is determined by the sequential \texttt{uv\_async\_file} event that precedes the \texttt{uv\_submit} event. Both events are always sequential(label $5$). The thread ID is used to match kernel events(labels $8$, $8b$). The SHT is then updated by creating the child entries corresponding to the detected kernel events. Figure~\ref{fig:exspan} depicts the resulting vertical span from the trace in Figure~\ref{fig:trace} as outputted by the analysis.

\section{Use-Cases}
\label{sec:use-cases}

This section investigates five realistic issues to illustrate the range of applications of the proposed new algorithms and methodology. Additionally, the use-cases demonstrate the effectiveness of the methodology in identifying and addressing performance issues for various problem types and contexts.

\par Use case $1$ represents a common scenario encountered by node.js application developers. We demonstrate the efficacy of Node Compass in detecting Regular Expressions Denial of Service (ReDOS) in application modules. Such a vulnerability permits malicious code to obstruct the EL in a particular phase, thereby delaying the execution of pending callbacks\citep{michaelgok} \citep{elblock}.
Use case $2$ represents one particular characteristic of our tool. It demonstrates its distinctiveness by revealing to what extent garbage collection operations can affect the execution time of an atomic operation in \texttt{Node.js}. Our tool is subjected to a realistic issue raised by developers on \texttt{Github}.
Use case $2$ represents one particular characteristic of our tool. It demonstrates the distinctiveness of our tool by revealing to what extent garbage collection operations can affect the execution time of an atomic operation in \texttt{Node.js}. Our tool is subjected to a realistic issue raised by developers on \texttt{Github}. 

The third use case illustrates how our tool can be used to detect memory leaks in Node.js applications. The performance is monitored by the utilisation of offline profiling of memory consumption, based on the collected trace.
Use case $4$ demonstrates how our tool can track Inter-process Communication (IPC) to conduct a performance analysis.

The last use case is a particular instance on how a root cause analysis can be driven by Node Compass.

The experiments were conducted with Node.js versions 12.22.3 and 16.1.0, on an I7 Core running Ubuntu 20.4 with 16 GB RAM. 

\subsection{Detecting REDOS Vulnerabilty in Node.js}

Our tool capability to spot ReDOS vulnerabilities in Node.js is highlighted in the first use case. It is not uncommon for certain modules loaded in applications to have this type of vulnerability. They have been investigated by \citet{staicu2018freezing}. Executing vulnerable code in the application can cause the event loop to block and errors to propagate to requests awaiting execution. Collected  high-level information from distributed tracers cannot expose the source of the problem. Accurately identifying the layer responsible for latency propagation requires a multilevel analysis approach. 

In order to demonstrate the effectiveness of our technique in identifying ReDOS vulnerabilities, we employed the benchmark resulting from the work of \citet{staicu2018freezing}. The latter encompasses numerous \texttt{JavaScript} RedOS vulnerabilities and exploits, that have been documented and detected across multiple accessible modules.
Node Compass was able to identify all performance issues caused by them. To demonstrate how our tool operates, we have selected one vulnerable module from the benchmark as a showcase.

\begin{figure*}[!ht]
\centering{\includegraphics[width=13.5cm,height=6cm]{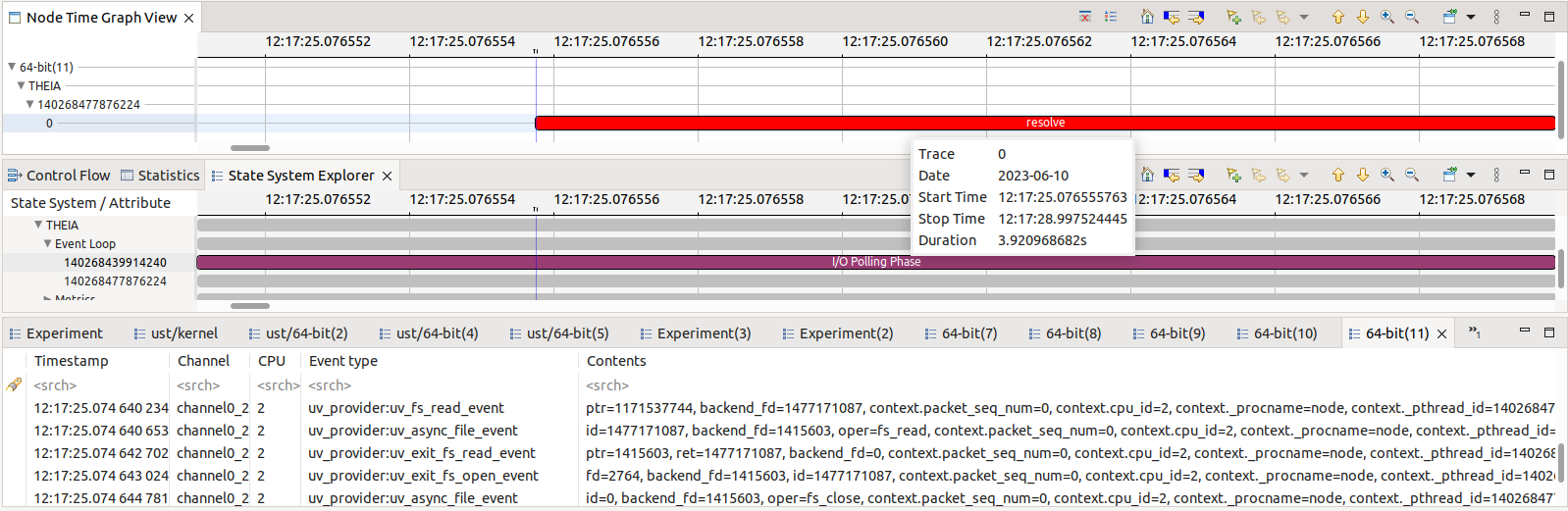}} 
\caption{Calling the resolve function from the DNS module. The event loop is stalled for 4 seconds, due to a REDOS vulnerabilty exploited. It takes 3.92 seconds for the resolve function to complete.}
\label{fig:regex}
\end{figure*}

Figure \ref{fig:regex} illustrates a performance issue that is triggered by calling the \texttt{resolve} \texttt{JavaScript} function of the Domain Name Service (DNS) \texttt{Node.js} module. It takes approximately $3.9$ seconds to execute. Our tool permits the correlation of high-level information with lower layers, such as the \texttt{libuv} and \texttt{V8} layers, in order to identify the underlying cause of the problem. Examining Figure \ref{fig:regex} reveals that the EL is halted in the polling phase, following the initiation of high-level \texttt{JavaScript} function execution. As a result of the developed analyses, the correlation between distinct layers is obvious. The activation of the tracepoint at the VM, by invoking the \texttt{exec} function , confirms the suspicion of a ReDOS exploit. The latter function processes regular expressions. The correlation of those informations, from different layers, results in the visual alignment of the events and the reconstruction of their internal operations.

\subsection{Correlating Performance issues with Garbage Collector Operations }

The \texttt{Node.js} runtime executes submitted \texttt{JavaScript} code at the interpreter level. An executed \texttt{JavaScript} function may exhibit latency that deviates from its normal behavior. Although the performance issue can be diagnosed in this case, the root cause cannot be identified. The vertical reconstruction of the execution sequences of a function enables the association of the different atomic operations with their execution context. To this extent, it allows the identification of atomic operations at the root of the performance problem, as well as the layer in which the issue is located.

Our approach effectiveness, and level of granularity, in exposing performance issues is demonstrated in this use case.  \texttt{Node.js} developers reported issue \#37583 on GitHub. A performance degradation was observed by using the \texttt{fs.promise.readFile} function, in comparison to the \texttt{fs.readFile}.

The performance issue was replicated in \texttt{Node.js} version $12$ by executing the suggested benchmark, which measures the performance of \texttt{fs.readFileSync, fs.readFile}, and \texttt{fs.promises.readFile}. Calling \texttt{fs.promise.readFile} to read a $1$ MB file was slower than \texttt{fs.readFile}. 

\begin{figure*}[!ht]
\centering{\includegraphics[width=13.5cm,height=5cm]{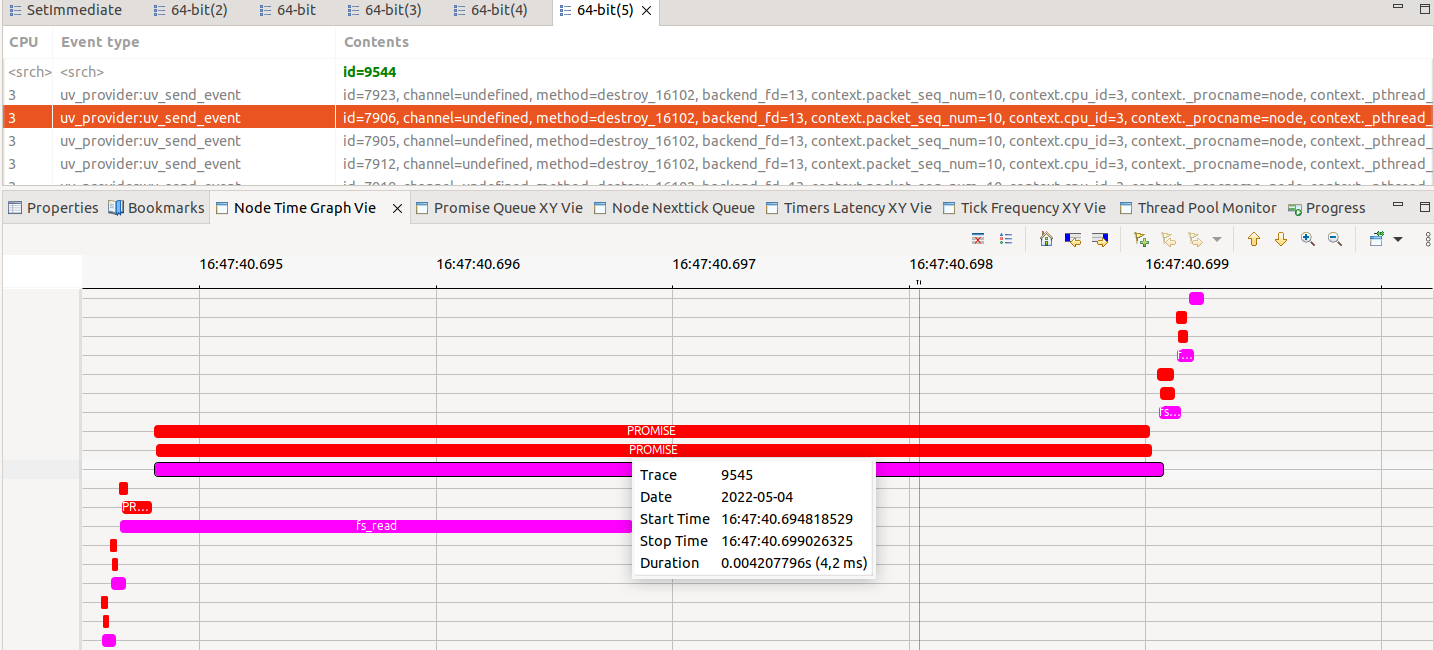}} 
\caption{VSM of the "fs.promise.readFile" execution. Atomic operations at \texttt{libuv} layer can been seen in the bottom perspective. The read operation is split into many atomic \texttt{fs\_read}. The top perspective show linear events triggered as result of the GC operation.}
\label{fig:fspromise}
\end{figure*}

\begin{figure*}[!ht]
\centering{\includegraphics[width=14cm,height=4cm]{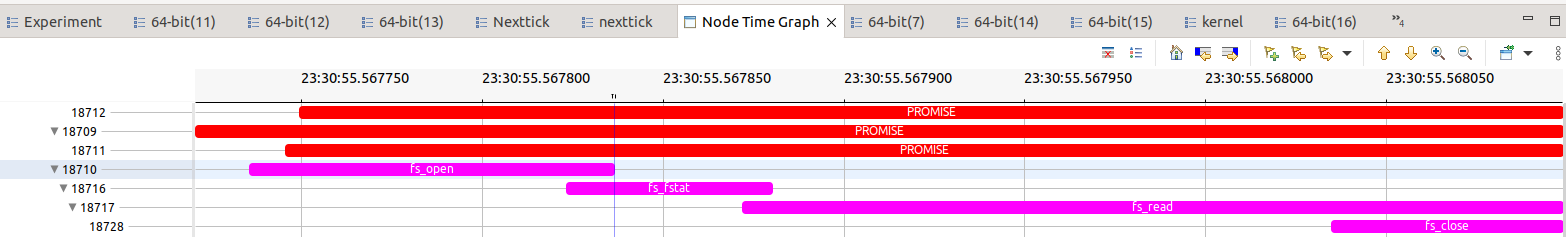}} 
\caption{VSM of the "fs.readFile" execution. Atomic operations at \texttt{libuv} layer can been seen. they are triggered by promises. To read a file, the atomic functions fs\_open, fs\_stat, fs\_read, fs\_close are called respectivly.}
\label{fig:fsread}
\end{figure*}

Our tool could exposed the execution flow of those two distinct reading functions. The \texttt{fs.readFile} function reads the entire file in a single function call. However, in the case of the \texttt{fs.promise.readFile}, the latter is split into many  \texttt{fs\_read} calls at the \texttt{libuv} layer. In Figure \ref{fig:fspromise}, we can see how memory operations from the garbage collector (GC) impact atomic read operations, increasing their latency by destroying memory resources linearly. Atomic read operations are blocked as a result. The general execution time for an atomic  \texttt{fs\_read} operation in Figure \ref{fig:fspromise} is approximately $50$ microseconds. The memory pass of the garbage collector disrupts certain \texttt{fs\_read} read operations that require between $3ms$  and $11ms$ to complete. This has the overall consequence of increasing the execution time of the higher level function, because promise type resources are created in the memory in large numbers, and must be destroyed by the GC to optimize memory usage.

\texttt{Node Compass} stands out for its high level of granularity, which enables the identification of performance issues with exceptional precision. Our introduced metric \texttt{ATTL} measures the execution time of an atomic operation. To the best of our knowledge, there is no performance analysis approach in the literature for accomplishing this in \texttt{Node.js}.

\subsection{Memory leak Detection}
In this use case, we demonstrate the effectiveness of our approach in profiling resources based on defined metrics. Node Compass analyzes GC behavior using $2$ specific metrics. The Time Spent in the Garbage Collection (TIGC) and the Time Between $2$ GC Operations (TBGC). We utilized the \texttt{Node.js} memory leaks benchmark \footnote{https://github.com/Joker666/nodejs-memory-leak}, which encompasses $4$ types: (Global, promise, cache, enclosures).
 
\begin{figure*}[!ht]
\centering{\includegraphics[width=13.5cm,height=3cm]{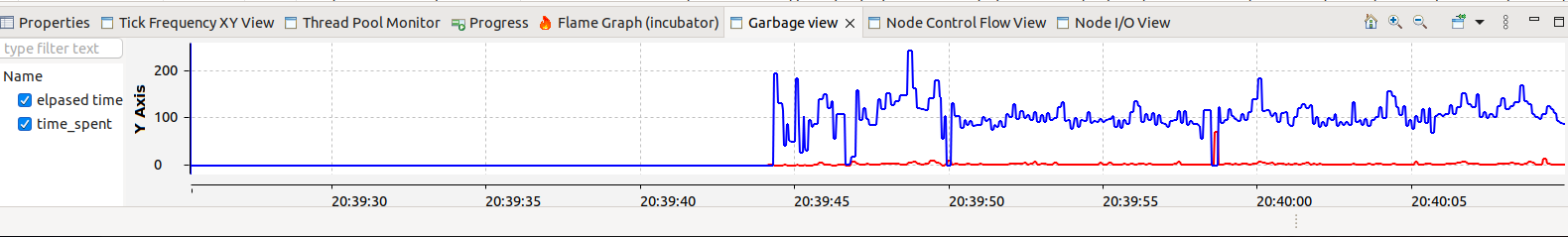}} 
\caption{Execution of the Cache type memory leak benchmark. Normal profile of the execution with no memory leak}
\label{fig:noleak}
\end{figure*}

Node Compass exposed memory leak-related performance issues suspicions in all $4$ cases using the defined metrics. We select one type of leaks (cache type) to discuss the Node Compass uncovering mechanisms, and demonstrate how our tool works. The same methodology applies to all types. The absence of memory leaks is reflected in Figure \ref{fig:noleak}, which depicts the regular execution profile of the Node.js process. 
The time spent between two garbage collections is constantly more than the time spent in garbage collection (red).

An optimal execution profile for the process application is achieved in such a situation. A difference can be observed in Figure \ref{fig:noleak} that shows the profile of the Node.js process exposing a memory leak. As seen, both metrics balance each other, raising a suspicion on heap problem. We observed this pattern in each memory leaks context.

\begin{figure*}[!ht]
\centering{\includegraphics[width=13.5cm,height=3cm]{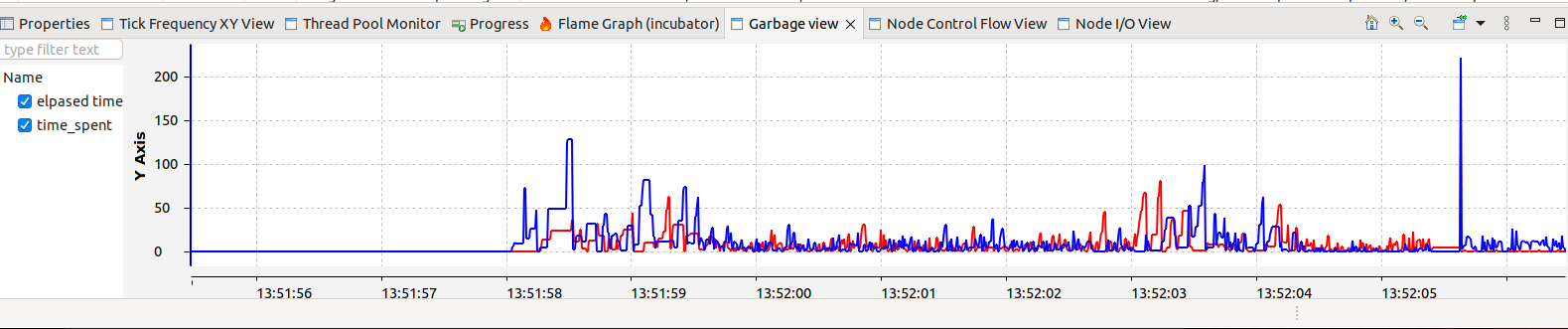}} 
\caption{Execution of the Cache type memory leak benchmark with memory leaks}
\label{fig:leak}
\end{figure*}

\subsection{Exposing Inter-process Communication issues}
\label{sec:interprocess}
We tested our performance analysis approach on a use case that involves inter-process communication (IPC). The latter was raised by the developers at \texttt{Ericson Canada}, the main committers of \texttt{Theia}. Our approach was successful in revealing the nature of the error. The \texttt{Node.js} \texttt{GitHub} site highlighted an issue with reference to \#43936, explaining that data on Windows gets "clogged" in extra pipes when forking a child\_process.

The distinguishing factor of this use case is its connection to the Windows NT environment. Our tool global application for multi-level performance analysis in Node.js is demonstrated in this context. Node Compass has the capability to trace, aggregate, and analyze events from the Windows OS without modifying its approach. Our instrumented \texttt{Node.js} versions can be compiled with a flag indicating the use of the "barectf" library in its trace collection architecture. The tracepoints within Node.js remain unchanged, with modifications made only at the tracer level. The tracer that was defined with \texttt{barectf} is responsible for collecting these events. \texttt{Barectf} produces a file in the Common Trace Format (CTF), the same format used by \texttt{LTTng}.

\begin{figure*}[!ht]
\centering{\includegraphics[width=13.5cm,height=3.5cm]{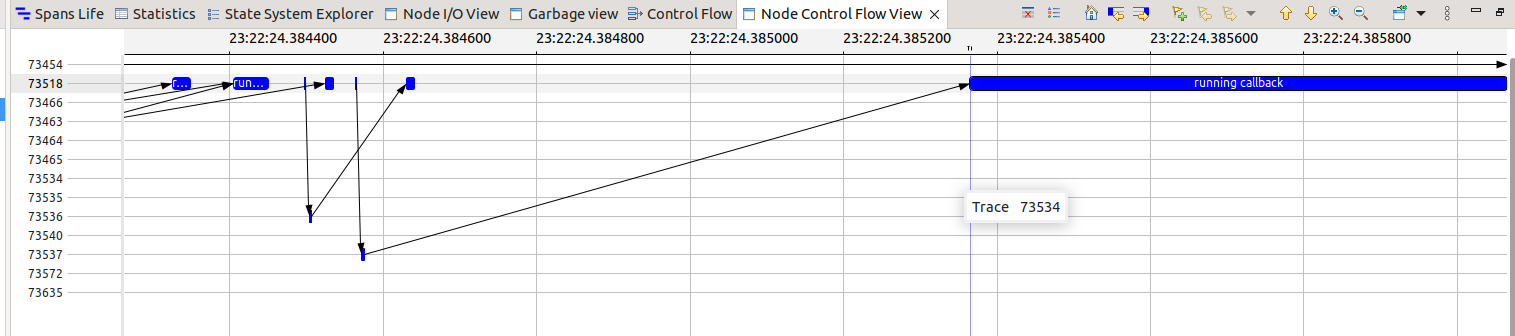}} 
\caption{Processes and threads interactions in Node.js. The arrows show the direction of the communications. The numbers are the threads and processes IDs}
\label{fig:ipc}
\end{figure*}
The use case pertains to a performance issue that occurs when a primary process (\texttt{main.js}) forks a child process (\text{fork.js}) with an additional open pipe. In normal operation, the child process automatically responds to pings from the parent process with pongs. In a similar manner, the parent process responds with a ping after reading the pong. This type of interaction makes process monitoring possible. Therefore, a rapid message exchange between the two processes is required.

During the execution of this procedure on Windows 10, the communication was hindered to a certain extent. Node Compass was used to identify the IPC communication flow within the ecosystem, to identify the problematic process, among others. Based on Figure \ref{fig:ipc} "Node Control Flow" view, the processes interactions can be observed, and the blocking processes can be identified. It is of utmost importance to explicitly indicate that the pipe is "overlapped", particularly in the context of \texttt{Windows}, to adhere to the Windows API requirements. The pipe for the various OS is not abstracted by Libuv. Therefore, the overlapped communication was denied, causing the parent process to remain blocked while it awaits the pong response from the child process.

\subsection{Root Cause Analysis}

In this subsection, we apply our technique to a major performance issue highlighted by \texttt{Theia} developers~\citep{theia}. It relates to issue 9514\footnote{\urlC}: The performance of \texttt{vscode.workspace.fs.readDirectory(Uri)} is a hundred times slower than VS Code, and issue 10684\footnote{https://github.com/acmeair/acmeair-nodejs}: Improve JSON-RPC communication performance. This specific issue has drawn much attention and interest, and has been the subject of intensive efforts. We were able to reproduce the performance issue locally, in order to evaluate our technique and show its usefulness and effectiveness.

\begin{figure}[!ht]
	\centering
	\includegraphics[width=8cm,height=4cm]{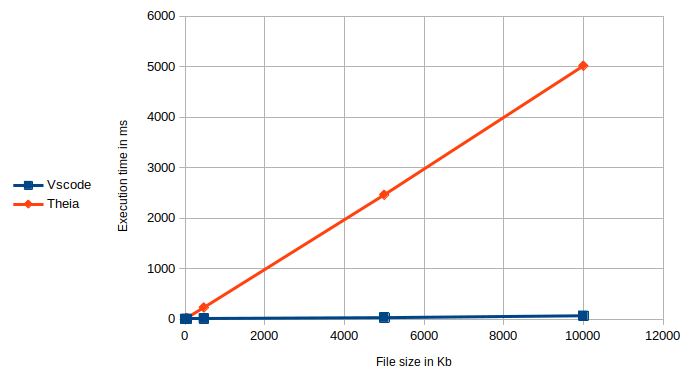}
	\caption{VSCode and Theia file reading average time.}
	\label{fig:graph1}
\end{figure}

To replicate the performance issue, we developed a VS Code plugin that calls the \texttt{fs.readFile} function to read a file. The plugin was explicitly tested in VS Code multiple times, with files of various sizes, and the average execution time was gathered. Then, we injected the plugin into \texttt{Theia} and ran the same tests. A significant performance issue was caused in \texttt{Theia} when the file size exceeded $450$ Kb. The compatibility of \texttt{Eclipse Theia}, with VS Code plugins and extensions, makes it a robust and adaptable framework, chosen by numerous developers. Eclipse \texttt{Theia} is at the core of several tools popular in industry.

\begin{figure*}[!ht]
	\centering
	\includegraphics[width=13.5cm,height=3cm]{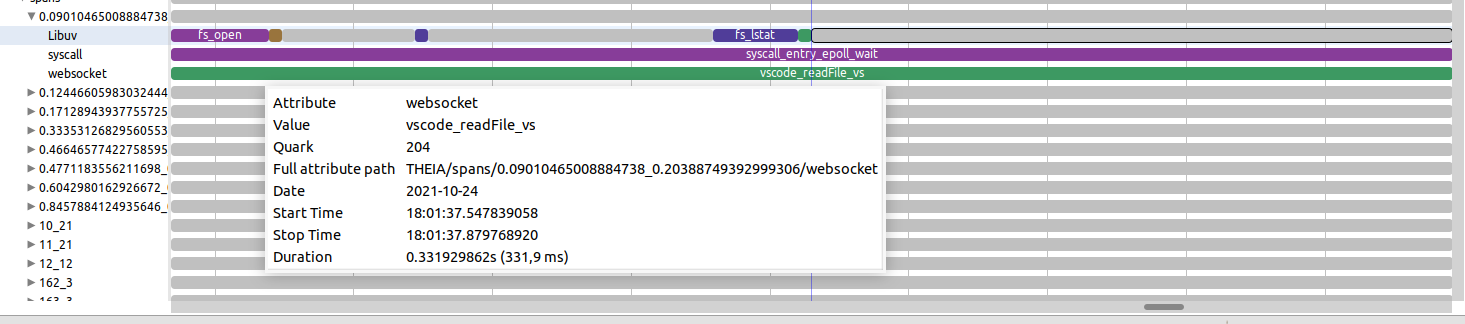}
	\caption{Plugin triggered vertical sequences. The sequences of \texttt{libuv} operations at the backend are presented along with the websocket span obtained from the execution of the function in the plugin. Syscall represents the sequences of system calls triggered by the plugin.}
	\label{fig:nodefull}
\end{figure*}

\begin{figure*}[!ht]
	\centering
	\includegraphics[width=14cm,height=4cm]{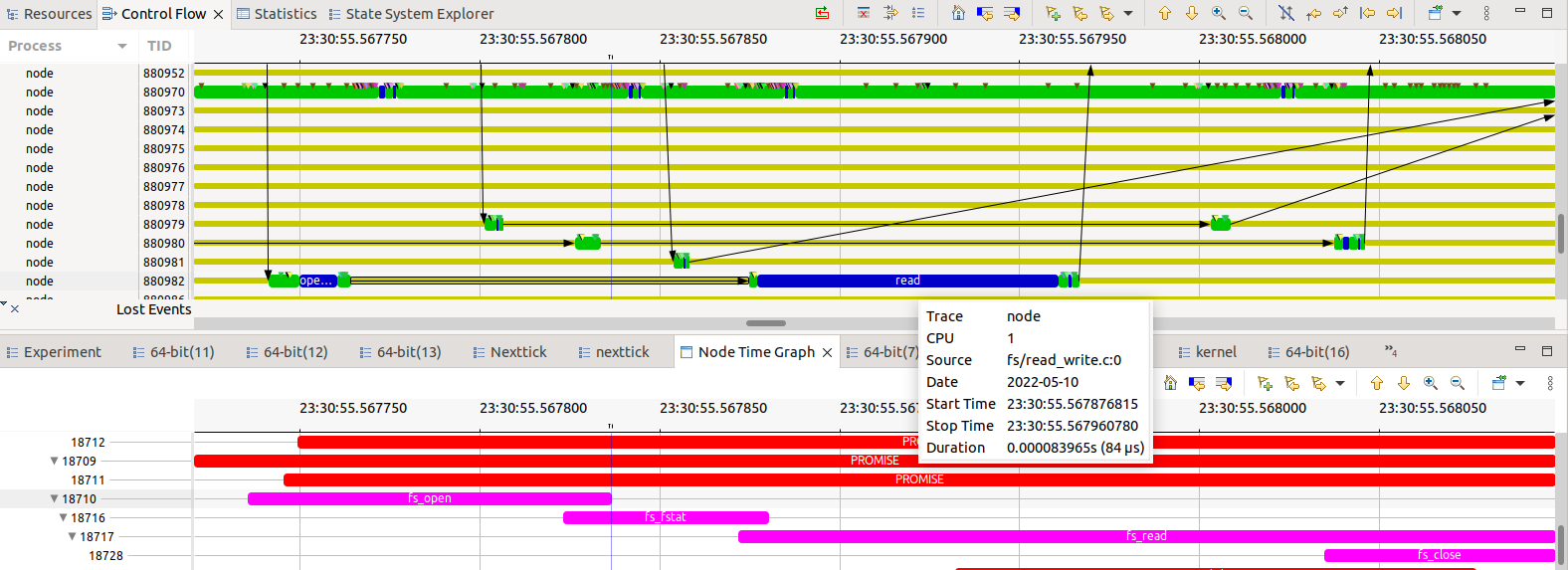}
	\caption{Vertical reading sequences of the Vscode plugin.}
	\label{fig:vscode1_good}
\end{figure*}

Figure~\ref{fig:graph1} depicts the outcomes of our experiments with multiple files of different sizes. We perceive that the processing time is acceptable for extremely small files, but it begins to deviate from those of VS Code when the file size exceeds 420 Kb. Beyond this, the performance issue is obvious. Node Compass was used to identify the root cause of the problem, and demonstrate its efficacy in resolving such challenges.

Under Theia, reading a 500 Kb file required approximately 331 ms. Figure~\ref{fig:nodefull} depicts a part of the Theia vertical execution flow. The latter includes information regarding the initial function \texttt{vscode\_readFile\_vs}, executed by the plugin to send the service information in the payload to the backend, via websocket and JSON-RPC. When the relevant Remote Procedure Call (RPC) service is invoked at the backend level, atomic operations are generated at the \texttt{libuv} level. Each I/O action in \texttt{libuv} triggers a system call. This information is displayed in Figure~\ref{fig:nodefull}, where we can also note the \texttt{vscode\_readFile\_vs} \texttt{readFile} method latency at around 331 milliseconds.

Clearly depicted in Figure~\ref{fig:nodefull}, the node process that runs the plugin enters the \texttt{epoll\_wait} mode after sending the request. Nonetheless, this does not account for the execution delay of the request. Digging further into the execution flow of the request in the backend, as depicted by the \texttt{Control Flow} perspective of Figure~\ref{fig:vscode1_good} on the top, it can be observed that the \texttt{read} system call requires only $84$ $\mu$s to perform. After executing the method, the backend process returns the response to the browser through Websocket. By narrowing the instrumentation to the Theia functions managing the communication layer, we observe that most of the request time is spent at the \texttt{JSON-RPC} communication layer. In this case, encoding binary data into a string is extremely inefficient in terms of performance. Mechanisms that allow sending direct binary data through Websockets should be implemented to handle such communications, instead of encoding it. Such an approach would improve the performance of the communication.

\section{Evaluation}
\label{sec:evaluation}

In this part, we measure the overhead associated with collecting the traces, and the time necessary to run our analyses. We also compare the overhead introduced by Node Compass with other \texttt{Node.js} tools. The primary purpose is to demonstrate a priori that our approach incurs a low overhead, when collecting data, in comparison to other \texttt{Node.js} analysis tools.

\begin{figure*}[!ht]%

    \centering
    \subfloat[\centering Services duration (without instrumentation)]{{\includegraphics[width=6cm, height=3cm]{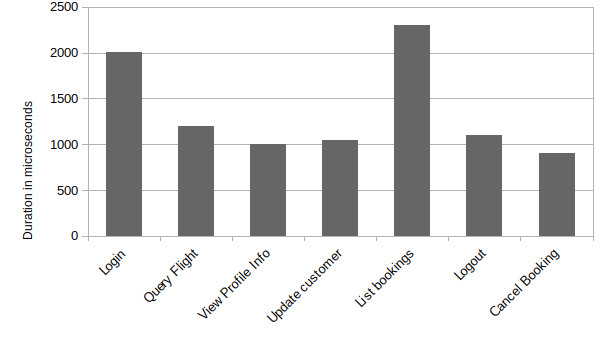} }}%
    \qquad
    \subfloat[\centering Services duration (with instrumentation)]{{\includegraphics[width=6cm, height=3cm]{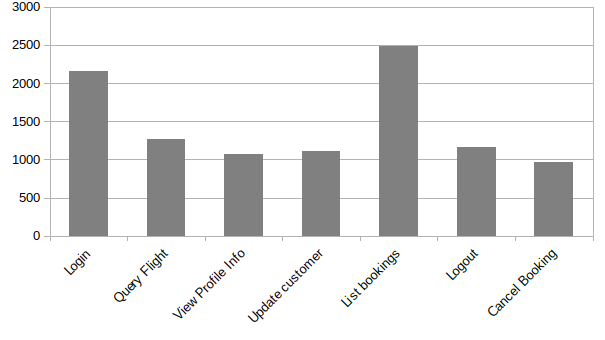} }}%
     \qquad
    \subfloat[\centering Async Hooks performances benchmarks.)]{{\includegraphics[width=7cm, height=3cm]{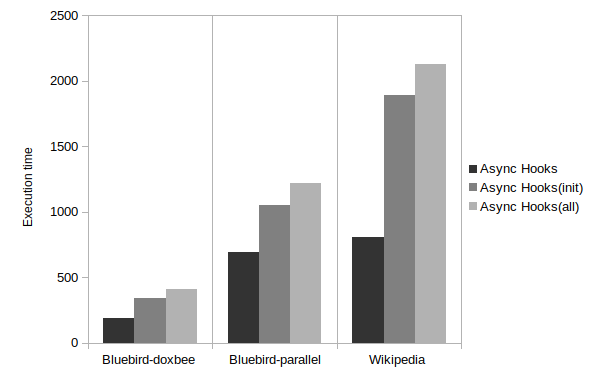} }}%
    \caption{Acmeair benchmark API paths. Time duration (Instrumented vs no instrumentation).}%
    \label{fig:acmeair-wiki}%
\end{figure*}

\begin{figure*}[!ht]%
    \centering
    \subfloat[\centering Acmeair benchmark CPU usage without instrumentation]{{\includegraphics[width=6cm, height=3cm]{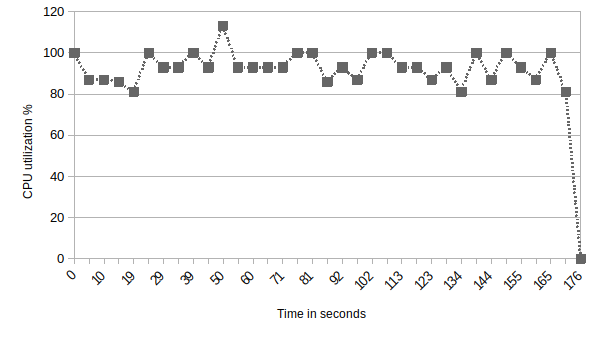} }}%
    \qquad
    \subfloat[\centering Acmeair benchmark memory usage without instrumentation]{{\includegraphics[width=6cm, height=3cm]{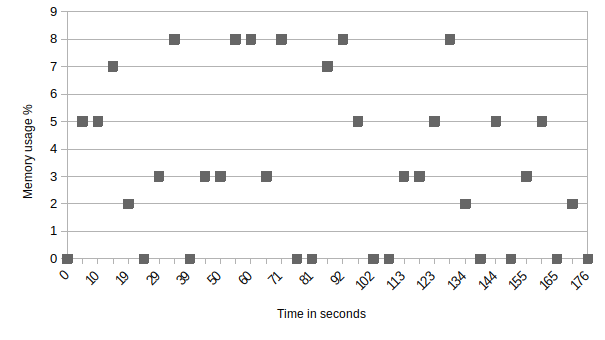} }}%
    \caption{Acmeair benchmark cpu and memory usage.}%
    \label{fig:acmeair-cpu-mem}%
\end{figure*}
\begin{figure*}[!ht]%
    \centering
    \subfloat[\centering CPU usage with instrumented Acmeair benchmark ]{{\includegraphics[width=6cm, height=3cm]{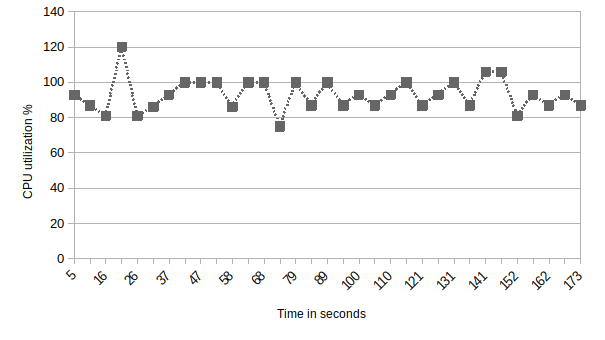} }}%
    \qquad
    \subfloat[\centering Memory consumption with instrumented Acmeair benchmark]{{\includegraphics[width=6cm, height=3cm]{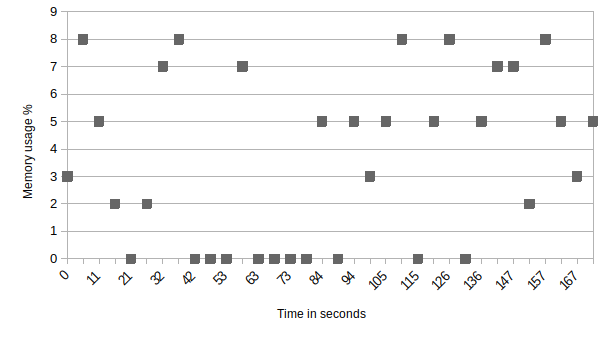} }}%
    \caption{Instrumented Acmeair benchmark cpu and memory usage.}%
    \label{fig:inst-acmeair-cpu-mem}%
\end{figure*}
Our tool overhead is compared to that of \texttt{NodeRacer}~\citep{endo2020noderacer} and \texttt{NodeAV}~\citep{chang2019detecting}. Although these two tools are used to solve specific problems, we will concentrate on their user trace collection architectures, which share with Node Compass the use of the \texttt{Async Hooks} API to monitor the lifecycle of asynchronous resources.

Figure \ref{fig:acmeair-wiki}(c) depicts the outcomes of the various micro-benchmarks we conducted to assess the overhead induced by the activation of \texttt{Async Hooks}. Observably, the activation of the \texttt{Init} callback imposes an very high cost on the performance. When a new asynchronous resource is created, this callback is initiated. As a result, it is possible to obtain information regarding the execution context of the resource. As can be observed, the situation deteriorates when all callbacks are activated, because it becomes necessary to monitor the resource initialization, execution start, execution end, and the destruction.  The challenge lies in the tricky nature of resource tracking at the top layer of \texttt{JavaScript}. Clearly, for each resource, the \texttt{JavaScript} and \texttt{C++} barrier must be crossed. This results in a significant increase in costs.

Unlike \texttt{Node Compass}, \texttt{NodeRacer} and \texttt{NodeAv} utilise \texttt{Async Hooks} in this manner. Unlike the other tools, Node Compass does not activate the \texttt{JavaScript} layer callbacks. Using the \texttt{Async Hooks} module, information about the execution context is obtained, whereas monitoring is performed at the \texttt{VM} level. The instrumentation of the latter with \texttt{LTTNg} enables the recovery of the execution context of the resources at the lowest level, resulting in a cost on the order of 8\% for our method. To the best of our knowledge, there is no multi-layered performance analysis approach for \texttt{Node.js} that incurs a comparable low cost. The experimental version of Node Compass is available for free download on GitHub\footnote{\urlA}.

\texttt{Acmeair}\footnote{https://github.com/acmeair/acmeair-nodejs}, a well-known benchmark for flight management applications, has also been used to test Node Compass. \texttt{Apache Jmeter} was used to produce the user load. The results of $2000$ requests executed with and without Node Compass instrumentation are displayed in Figure \ref{fig:acmeair-wiki}. As can be seen, the additional cost incurred is acceptable.

Figure~\ref{fig:acmeair-cpu-mem} and Figure~\ref{fig:inst-acmeair-cpu-mem} illustrate resources consumption during the benchmark execution. Respectively, with no instrumentation and with instrumentation. We observe that the differences between the two cases are not discernible in terms of memory and especially processor usage.

\begin{figure*}[!ht]%
    \centering
    \subfloat[\centering CPU consumption during trace loading and analysis]{{\includegraphics[width=6cm, height=3cm]{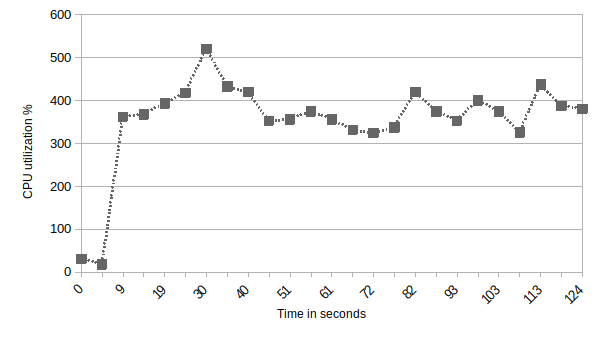} }}%
    \qquad
    \subfloat[\centering Memory consumption during trace loading and analysis]{{\includegraphics[width=6cm, height=3cm]{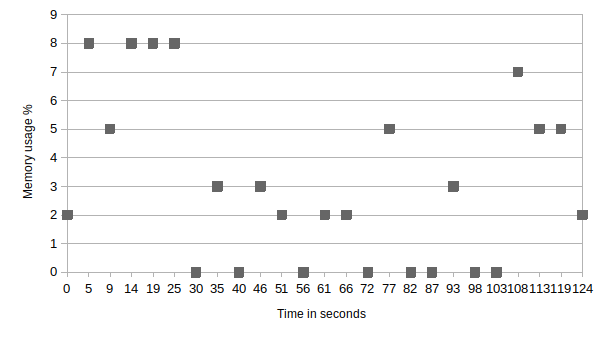} }}%
    \qquad
    \subfloat[\centering Number of threads over running time]{{\includegraphics[width=6cm, height=3cm]{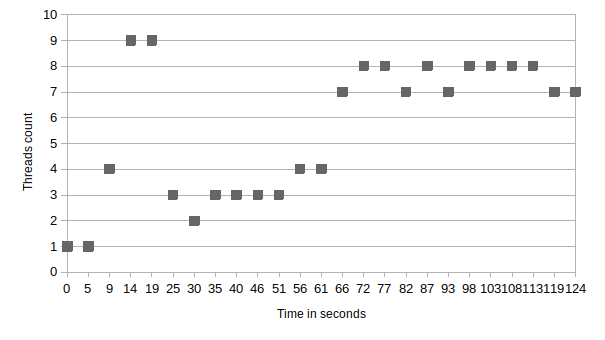} }}%
    \qquad
    \subfloat[\centering Trace size vs analysis running time]{{\includegraphics[width=6cm, height=3cm]{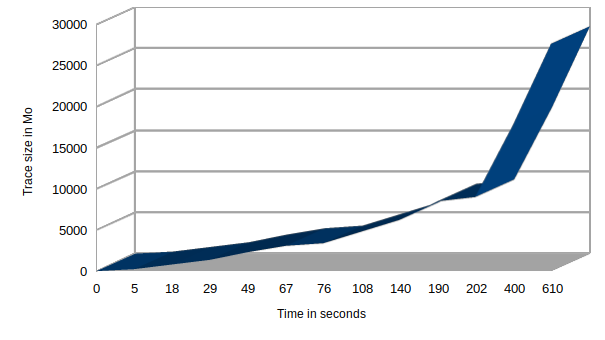} }}%
    \caption{Trace loading and analysis with Node Compass.}%
    \label{fig:java-cpu-mem}%
\end{figure*}

We evaluated the cost of conducting our analyses for various trace sizes. Figure~\ref{fig:java-cpu-mem} depicts the utilisation of the CPU (a) and memory (b) during the loading and execution of the various trace-related analyses. We can observe the distribution of intervening threads over time, given the concurrent nature of the analysis execution.
As the execution time of the analysis depends on the size of the collected trace, Figure~\ref{fig:java-cpu-mem} (d) displays the results of the latter in the context of different sizes. Instrumenting a \texttt{Node.js} application with Node Compass allows to collect a trace over time. The ratings provided represent the worst-case scenario. It is not necessary to obtain a kernel trace for each analysis. Some analyses do not require their use and are limited to collecting a user trace from the internal \texttt{Node.js} layers. Analyses such as I/O processing rely on kernel events. The size of a user trace is approximately $450$ times smaller, in comparison to a kernel trace, when all events are collected.
\section{Conclusion}
\label{sec:conclusion}

\texttt{JavaScript} is an environment with great potential and is the basis for thousands of applications currently in production worldwide. It enjoys widespread popularity among developers~\citep{shah2017node, tuto, zhu2018scalability}. Its flexibility and continuously expanding ecosystem make it one of the most used distributed application environments. Understanding the performance of \texttt{JavaScript} applications is crucial for developers. Therefore, tools and adequate approaches are necessary to help detect the various problems at the basis of performance degradations.

We introduced a novel technique based on the vertical reconstruction of the execution flow of \texttt{Node.js} applications. Vertical profiling and calling-context profiling are effectively attained through the application of two distinct techniques. Through the identification of execution patterns in the trace, our method correlates events between the various layers. The end result is an interactive interface for identifying potential performance issues.

Our technique enables performance analysis from several angles. It is a global approach that sees the system as being complex and dynamic, responding to a certain number of expectations. Therefore, unlike most tools that address specific problems, such as atomicity violations, detection of race issues, or that rely on higher level operations latency, to diagnose performance problems, Node Compass adopts a multi-level approach. This enables the reconstruction of the execution sequences of a task. In this context, depending on the conducted analysis, the vast majority of performance issues can be pinpointed, each in the responsible layer. The reconstruction of the application execution flow is an essential element in the granularity of the approach, in order to detect the problem root cause. One exciting area to consider for future work is to perform a critical path analysis of the different requests under \texttt{JavaScript}, in a distributed system context. It would allow a defined execution model, and help reveal bottlenecks related to specific requests processing and I/O threads management.
\section{Acknowledgments}
\label{sec:ack}
We would like to gratefully acknowledge the Natural Sciences and Engineering Research Council of Canada (NSERC), Prompt, Ericsson, Ciena, AMD and EfficiOS for funding this project.

\bibliography{cas-refs}



\end{document}